\RequirePackage{fix-cm}
\documentclass{svjour3}[10/04/2014]       % onecolumn (second format)
\usepackage{graphicx}
\usepackage{siunitx}
\usepackage[square, numbers]{natbib}
\usepackage[tight]{subfigure}
\usepackage{color}
\usepackage{verbatim}
\usepackage{hyperref}
\journalname{Experimental astronomy}
\newcommand\aap{A\&A}%
\graphicspath{{./figures/}}
\date{10/4/2014}
\begin{document}
\title{The effect of sideband ratio on line intensity for Herschel/HIFI}


\author{Ronan Higgins         \and
        David Teyssier            \and
        Colin Borys          \and
        Jonathan Braine      \and
        Claudia Comito      \and
        Bertrand Delforge   \and
        Frank Helmich       \and
        Michael Olberg	       \and
        Volker Ossenkopf     \and
        John Pearson \and
        Russell Shipman
}


\institute{R. Higgins, C. Comito and V. Ossenkopf \at
              I. Physikalisches Institut\\
              Universit\"{a}t zu K\"{o}ln\\
              Z\"{u}lpicher Straße 77\\
              50937 K\"{o}ln, Germany \\
              Tel.:  +49 221 4701707
              \email{higgins@ph1.uni-koeln.de} %  \\
           \and
           D. Teyssier \at
              Herschel Science Centre\\
              European Space Astronomy Centre (ESAC)\\
              P.O. Box 78, 28691 Villanueva de la Ca\~{n}ada\\
              Madrid, Spain
           \and
           C. Borys \at
           California Institute of Technology\\
           1200 E California Blvd.\\
           MS 100-22\\
           Pasadena, CA, 91125, USA\\
           \and
           J. Braine \at
          Observatoire de Bordeaux, LAB \\
          2 rue de l'Observatoire\\
          BP 89, 33270 Floirac, France
           \and
           B. Delforge, F. P. Helmich and R. Shipman\at
          SRON Netherlands Institute for Space Research \& Kapteyn Astronomical Institute\\
          University of Groningen\\
          PO-Box 800, 9700 AV Groningen\\
          The Netherlands
           \and
           M. Olberg \at
          Onsala Space Observatory\\
          Chalmers University of Technology\\
          S-43992 Onsala, Sweden
           \and
           J. C. Pearson \at
          Mail Stop 301-429, Jet Propulsion Laboratory \\
          4800 Oak Grove Drive\\
          Pasadena, CA 91109, USA     
          }

\date{Received: date / Accepted: date}

\maketitle

\begin{abstract}
The Heterodyne Instrument for the Far Infrared (HIFI) on board the Herschel Space Observatory is composed of a set of fourteen double sideband mixers. We discuss the general problem of the sideband ratio (SBR) determination and the impact of an imbalanced sideband ratio on the line calibration in double sideband heterodyne receivers.  The HIFI SBR is determined from a combination of data taken during pre-launch gas cell tests and in-flight. The results and some of the calibration artefacts discovered in the gas cell test data are presented here along with some examples of how these effects appear in science data taken in orbit.

\end{abstract}

\section{Introduction}
\label{intro}

The Herschel Space Observatory (\cite{pilbratt2010}) was launched on May 14th 2009 and successfully observed objects in the Sub-millimetre and Far Infrared (FIR) bands from the Solar System to the most distant reaches of the Universe.  
%The boiling off of the Helium coolant ended the mission on April 29$^{\rm th}$ 2013. 
The mission ended on April 29$^{\rm th}$ 2013 when the Helium coolant boiled off. Its spectral coverage was provided by three instruments, PACS, SPIRE and HIFI. PACS (Photodetecting Array Camera and Spectrometer) was an imaging camera and low-resolution integral field spectrometer covering wavelengths from 55 to 210\,\si{\micro\metre} \cite{poglitsch2010}. SPIRE (Spectral and Photometric Imaging Receiver) was also a camera and a Fourier transform spectrometer (FTS) covering wavelengths from 194 to 671\,\si{\micro\metre} \cite{griffin2010}. HIFI (Heterodyne Instrument for the Far-Infra-red) was a heterodyne detector and spectrometer providing high resolution spectroscopy capability over two continuous frequency ranges of 488--1272 and 1430--1902\,GHz \cite{deGraauw2010}. 
This work addresses the calibration of the HIFI instrument.
\par 
HIFI covers its spectral range using 14 heterodyne detectors, mixing down the FIR signal to radio frequencies. They are organised in 7 bands with 2 mixers each, sensitive to orthogonal polarizations.  The mixers in each band are pumped by a pair of Local Oscillator (LO) chains, covering respectively the lower (LO chain {\it a}) and upper (LO chain {\it b}) frequencies of a band tuning range. Table \ref{mixer_overview} provides an overview of the frequency coverage and mixer technologies used in HIFI. 
\par

\begin{table}[!t]
\centering 
%\begin{tabular}{|p{3.8cm}|p{2.3cm}|p{2.9cm}|p{2.5cm}|p{3.3cm}|}
%\begin{tabular}{|p{0.55cm}|p{1.6cm}|p{2.2cm}|p{2.1cm}|p{1.8cm}|p{2.4cm}|}
\begin{tabular}{|l|l|p{1cm}|l|l|l|l|}
\hline
Mixer      & LO Frequency   & IF BW & Detector                      & Beam      & Feed and coupling\\
band       & range (GHz)& (GHz) & technology                   & combiner    & structure\\\hline
\textbf{1} & 488--628    & 4--8 & SIS$^a$        \cite{delorme2005}                    & Beamsplitter& corrugated horn\\
           &            &     &    & microstrip            & and waveguide\\\hline
\textbf{2} & 634--794    & 4--8 & SIS$^a$     \cite{teipen2005}                   & Beamsplitter&corrugated horn\\
           &            &     &    & microstrip                 &and waveguide\\\hline
\textbf{3} & 807--953    & 4--8 & SIS$^a$       \cite{delange2003}                   & Diplexer&corrugated horn\\
           &            &     &    & microstrip          &and waveguide\\\hline
\textbf{4} & 957--1114   & 4--8 & SIS$^a$      \cite{delange2003}                     & Diplexer&corrugated horn\\
           &            &     &   & microstrip              &and waveguide\\\hline
\textbf{5} & 1116--1272  & 4--8 & SIS$^b$              \cite{karpov2007}               & Beamsplitter&lens and twin slot\\
           &            &     &             & microstrip            & planar antenna\\\hline
\textbf{6} & 1430--1698  & 2.4--4.8 & HEB$^c$        \cite{cherednichenko2008}                     & Diplexer&lens and twin slot\\
           &            &         &            &                       & planar antenna\\\hline
\textbf{7} & 1701--1902  & 2.4--4.8 & HEB$^c$       \cite{cherednichenko2008}                            & Diplexer &lens and twin slot\\
           &            &         &            &                         & planar antenna\\\hline

\end{tabular}
\caption{Overview of mixer technology, materials and implemented antenna technology \cite{deGraauw2010}. $^a$ Nb-Al$_{2}$O$_{3}$-Nb, $^b$ Nb-AlN-NbTiN, $^c$ NbN Phonon cooled.}
\label{mixer_overview}
\end{table}

\par 
Observing in the environment of space allows unobstructed by atmosphere coverage of the entire HIFI frequency range. It also removes the day night cycle constraints and allows for continuous visibility of a source over an observational day. From a calibration perspective, observing from space also gives a number of advantages over ground based telescopes. The absence of the atmosphere from the calibration equation removes one of the major sources of error in flux determination \cite{guan2012}. Owing to a careful thermal control of the detection chain, a high temperature stability is achieved, that reduces mixer gain variation, and returns better data quality with less baseline features such as standing waves and baseline distortions. However, at the shorter wavelengths of the Hot Electron Bolometer (HEB) mixers, data quality was at times poor due to system instability \cite{higgins2009, kooi2009}. 
\par
In anticipation of this unique environment ambitious calibration accuracies were sought, with a baseline calibration uncertainty of 10\% and a goal calibration uncertainty of 3\% \cite{roelfsema2012}. The main sources of calibration error inherent to a double sideband heterodyne system were determined to be the sideband ratio (hereafter SBR), standing waves, and the calibration load coupling. Specific tests were implemented prior to launch to constrain these error sources. They are described in \cite{teyssier2008}.
\par 
In this paper we discuss the sideband ratio derived from flight data and from the pre-launch test campaign conducted with a gas cell. Section \ref{side_band_ratio} gives the background on the significance of the sideband ratio and Section \ref{sbr_imbalance} describes the various phenomena involved in defining its characteristics. Section \ref{gascell_sideband_ratio} summarizes the results from the gas cell test campaign. Section \ref{side_band_ratio_in_data} provides examples of how the SBR manifests in certain areas of the HIFI frequency range.
A discussion section follows summarizing the lessons learned from this calibration effort.

\section{What is the sideband ratio and why is it important}
\label{side_band_ratio}
The HIFI mixers are double sideband (DSB) mixers which detect signals in an upper (USB) and lower (LSB) sideband simultaneously as the mixing process does not distinguish between positive and negative differences between sky frequency and LO frequency. The width of this sideband is the mixer IF (intermediate frequency) bandwidth (BW) which is dependent on the mixer design, see Table \ref{mixer_overview}. The detected signal is calibrated using the two-load calibration  \cite{ossenkopf2003,magnum2002}. This method requires the observation of a reference blank sky position, together with internal hot and cold sources. Since the temperature of the hot and cold loads and their coupling to the mixers are known, one can assign an intensity to the sky signal. Using this method one can determine the double sideband intensity of the detected signal. To accurately assign an intensity to a given spectral channel in a given sideband one must have an \textit{a priori} knowledge of the fraction of the double sideband signal that belongs to the respective upper and lower sideband at a given LO frequency. This fraction is known as the {\it sideband ratio}. 
\par 
The sideband ratio, $R$, is the ratio of one sideband gain over the other and defined as:
\begin{equation}
\label{side_band_ration_formula}
R=\frac{\gamma_{usb}}{\gamma_{lsb}}
\end{equation}
where $\gamma_{usb}$ is the response in the USB and $\gamma_{lsb}$ is the response in the LSB. The total instrument response, $\gamma_{dsb}$, is then:
\begin{equation}
\label{dsb_total}
\gamma_{dsb}=\gamma_{usb} + \gamma_{lsb}
\end{equation}
In the HIFI data processing pipeline a dedicated pipeline step {\tt doSidebandGain} converts the double sideband intensity into separate upper and lower sideband intensities \cite{avruch2013}. In this step the DSB spectra is divided by the {\it normalized} sideband ratio, $G_{usb}$, to return the upper sideband intensity, and independently divided by the factor $1-G_{usb}$ to return the lower sideband intensity. The normalized upper sideband gain is related to the sideband ratio, $R$, as follows:
\begin{equation}
\label{g_ssb_equation}
G_{usb} =\frac{\gamma_{usb}}{\gamma_{dsb}}=\frac{\gamma_{usb}}{\gamma_{usb} + \gamma_{lsb}}= \frac{R}{1+R}
\end{equation}
 $G_{usb}$ describes what fraction of the total double sideband response stems from the upper sideband. Then following from equation \ref{dsb_total}, $1-G_{usb}$ is the fraction of that DSB response that comes from the lower sideband\footnote{In case of LO impurities, providing coupling to multiple frequencies, the heterodyne receiver can also be considered as a multi-sideband receiver where the normalized gains of all sidebands add up to unity}.
\par 
In an ideal double sideband heterodyne receiver both sidebands have equal gain and hence $G_{usb}$ is 0.5. This is the base assumption for HIFI; however, as we will see in the following sections, the sideband ratio can vary considerably over a mixer band.

\section{Reasons for sideband ratio imbalance}
\label{sbr_imbalance}
\begin{figure}
\centering
\includegraphics[angle=-90.0,width=\textwidth]{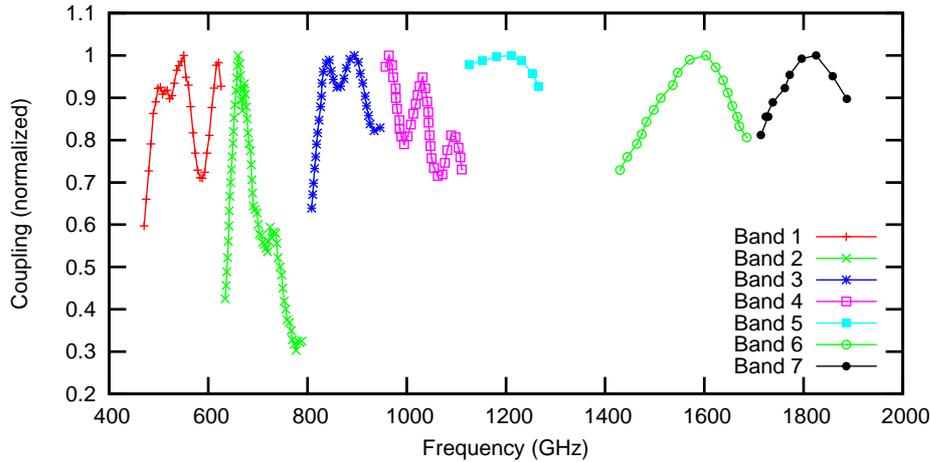}
\caption{Normalized mixer broadband coupling for bands 1--7 determined through FTS measurements with the mixer biased in direction detection mode. The data for band 1 are taken from \cite{delorme2005}, band 2 is from \cite{teipen2005}, bands 3 \& 4 are from \cite{delange2003}, band 5 is from \cite{karpov2007} and the data for bands 6 \& 7 are taken from \cite{cherednichenko2008}.}
\label{fts_coupling}
\end{figure}

\subsection{Mixer broadband coupling}
\label{fts}

%\subsubsection*{HIFI mixer broadband coupling measurement}

The degree of sideband imbalance is mainly driven by the variation of the mixer broadband coupling. This coupling was measured for all HIFI mixers prior to flight using an FTS. For this test, the FTS is coupled to the mixer and swept in order to extract the mixer output power at each frequency, and derive the mixer broadband coupling.
It should be noted, however, that the FTS measurements are taken with the mixer biased to be sensitive to direct detection and not in heterodyne mode (i.e. the mixer is not {\it pumped} by an LO source).
Fig. \ref{fts_coupling} provides an overview of the coupling measured for each HIFI mixer. It shows that mixers 1-4 have large variations in coupling compared to mixers 5-7. The reasons for this vary. Mixers 3 and 4 were measured with the same experimental setup and were both affected by a non optimum cryostat window which introduced a large standing wave into the measured data. Bands 1 and 2 were taken under better conditions and provide a clearer picture of this coupling.
Bands 5-7 show less frequency variation which is likely due to the coupling structure or mixer technology. 
%Bands 5-7 have a twin slot antenna while bands 1-4 have a wave guide antenna, see \ref{mixer_overview}. 
\par

\begin{figure}
\subfigure[3 LO frequencies (dashed lines) with their associated lower and upper sidebands shown in red and blue respectively on top of the band 2 mixer broadband coupling measurement.]{
\label{fts_plot_example}
\includegraphics[angle=-90.0,width=\textwidth]{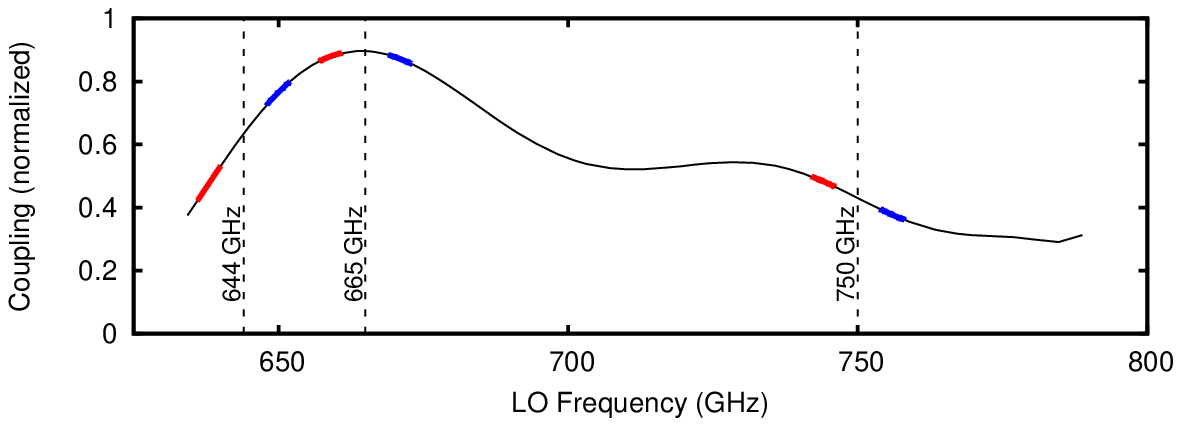}
}

\subfigure[Upper (blue) and lower (red) sideband mixer response at different LO frequencies plotted against IF.]{
\vspace{10.0in}
$
\begin{array}{ccc}
\includegraphics[angle=-90,width=0.33\textwidth]{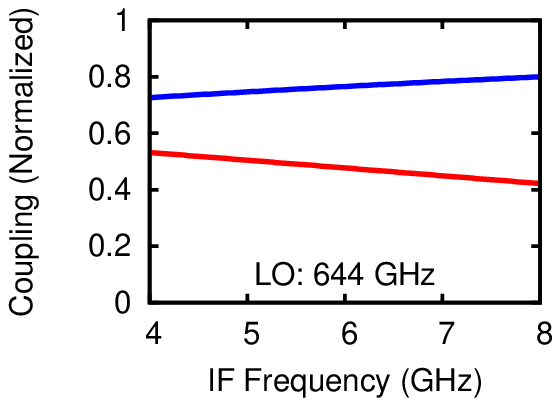}&
\includegraphics[angle=-90,width=0.33\textwidth]{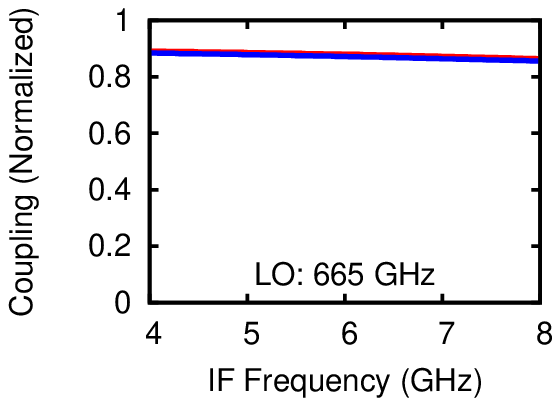}&
\includegraphics[angle=-90,width=0.33\textwidth]{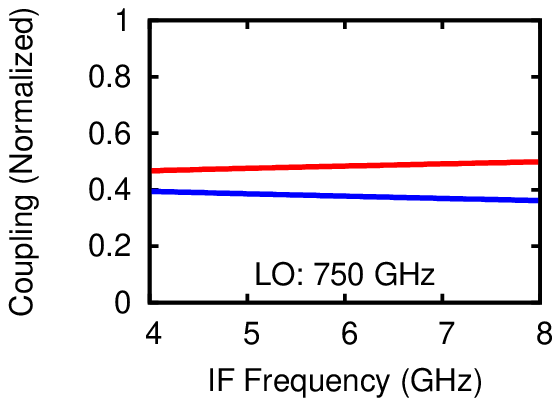}
\end{array}
$
}

\subfigure[Resulting sideband ratio, $G_{usb}$.]{
$
\begin{array}{ccc}
\includegraphics[angle=-90,width=0.33\textwidth]{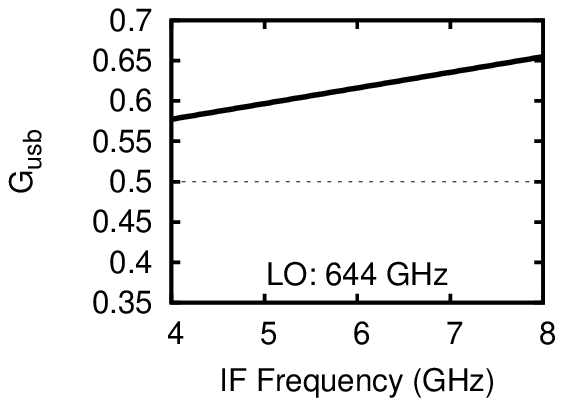}&
\includegraphics[angle=-90,width=0.33\textwidth]{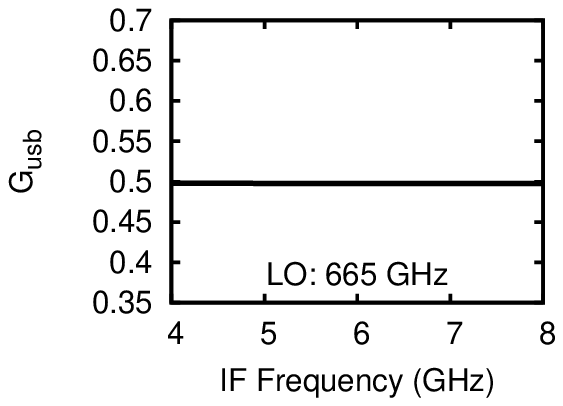}&
\includegraphics[angle=-90,width=0.33\textwidth]{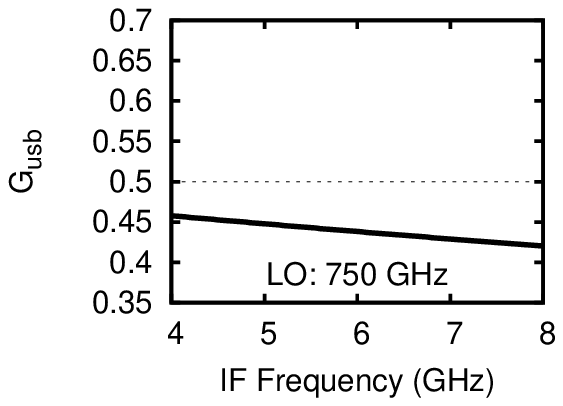}
\end{array}
$
\label{fts_sbr_slices}
}
\caption{Sideband ratio, $G_{usb}$, calculated from FTS broadband coupling measurements in band 2.}
\label{fts_sbr_example}
\end{figure}

The FTS spectra shown in Fig. \ref{fts_coupling} can be converted into a sideband gain plot using equation \ref{g_ssb_equation}. Fig. \ref{fts_sbr_example} shows an example of the variation in upper and lower sideband gain and the resulting sideband ratios in the HIFI band 2 for 3 different LO frequencies. The effect is most prominent in regions where there is a large slope in the broadband coupling. In these regions, such as the band edges, one sideband is dominant over the other. At the lower band edge e.g. the LSB coupling is less than the USB one and hence we have a normalized sideband ratio, $G_{usb}$, higher than 0.5. The opposite effect is seen at 750\,GHz where the USB coupling is weaker than that of the LSB and hence $G_{usb}$ is less than 0.5. In regions, such as at 665 GHz, where the coupling is equal in both sidebands we have a balanced sideband gain scenario, but still a small slope in the gain is seen over the IF bandwidth.
\par 
Returning to Fig. \ref{fts_coupling} one can expect that mixers 5--7 have less sideband ratio imbalance since the broadband coupling varies slowly with sky frequency compared with bands 1--4. Also from the Fig. \ref{fts_sbr_example} it should be noted that the sideband ratio is more extreme towards higher IF's since these regions are at the maximum frequency separation and so are more sensitive to slopes in broadband coupling. 
Since the HEB mixers have a lower IF than the SIS mixers, they are less sensitive to sideband ratio imbalances. This will be discussed further in section \ref{sect_heb}.
\par 
\begin{figure}
\centering
%  716546.5590   .0600 -2.4681 2  693.6672119 -60001 10159          58           
%  728654.2827   .0175 -2.4965 2  717.5687121  60001 10160          59 
\subfigure[Simulated OCS spectrum shown between 712 and 730\,GHz. The resulting upper and lower sidebands components are highlighted in blue and red respectively when observing with an LO frequency of 721.3 GHz and an IF bandwidth from 4 to 8\,GHz.]{
\label{sbr_line_examples_ssb_spectra}
\includegraphics[angle=-90,width=\textwidth]{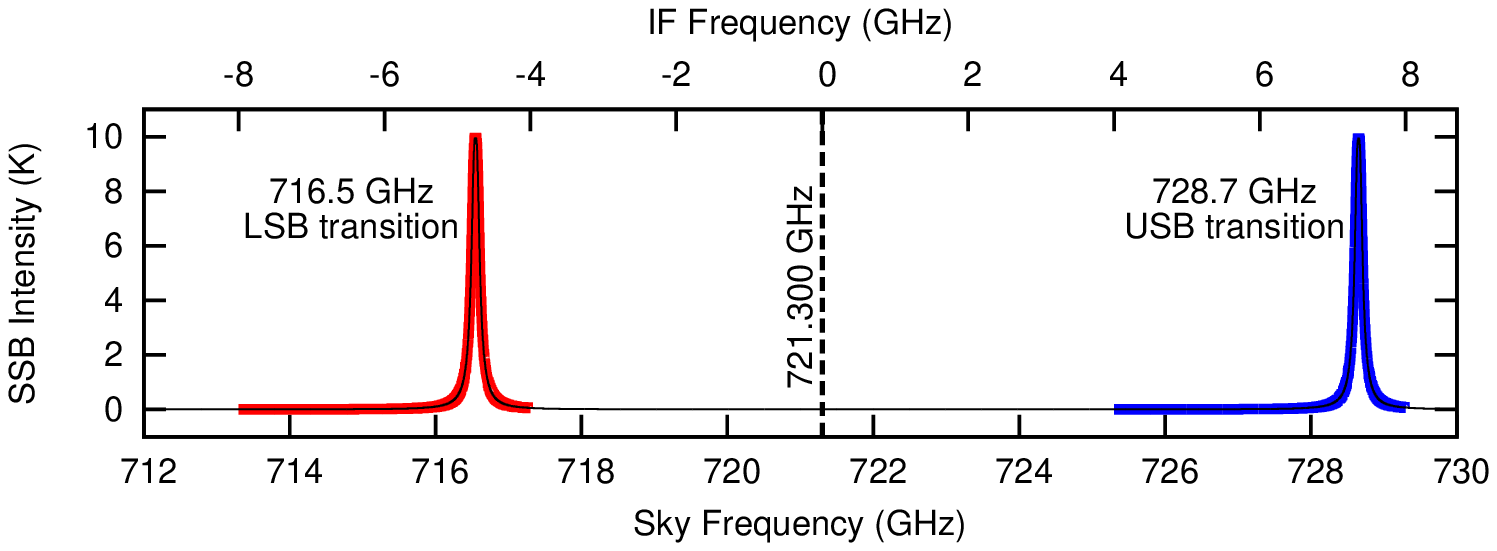}
}

\subfigure[Simulated HIFI double sideband spectrum applying the USB-dominated sideband ratio at 644 GHz from Fig. \ref{sbr_line_examples_gssb} (red) and a balanced sideband ratio scenario (green).]
{
\label{sbr_line_examples_dsb}
\includegraphics[angle=-90,width=\textwidth]{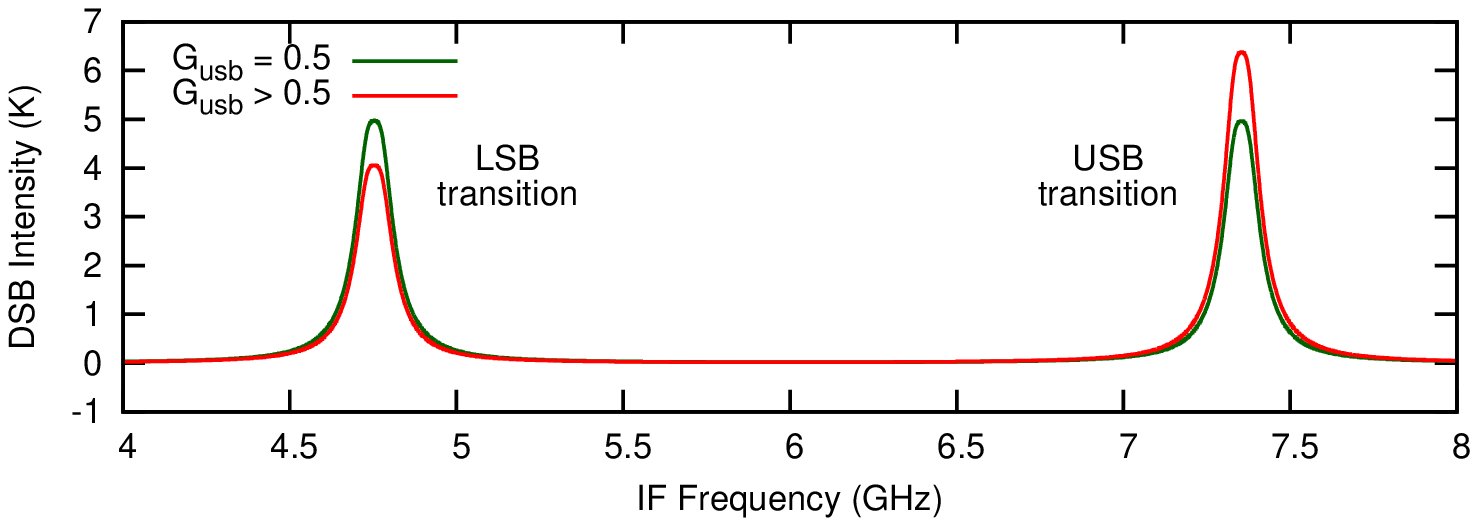}
}

\subfigure[Sideband ratio versus IF at 644 GHz from Fig. \ref{fts_sbr_slices} (red) compared with a sideband ratio balanced scenario where $G_{usb}$ is 0.5 across the IF band (green).]
{
\label{sbr_line_examples_gssb}
\includegraphics[angle=-90,width=\textwidth]{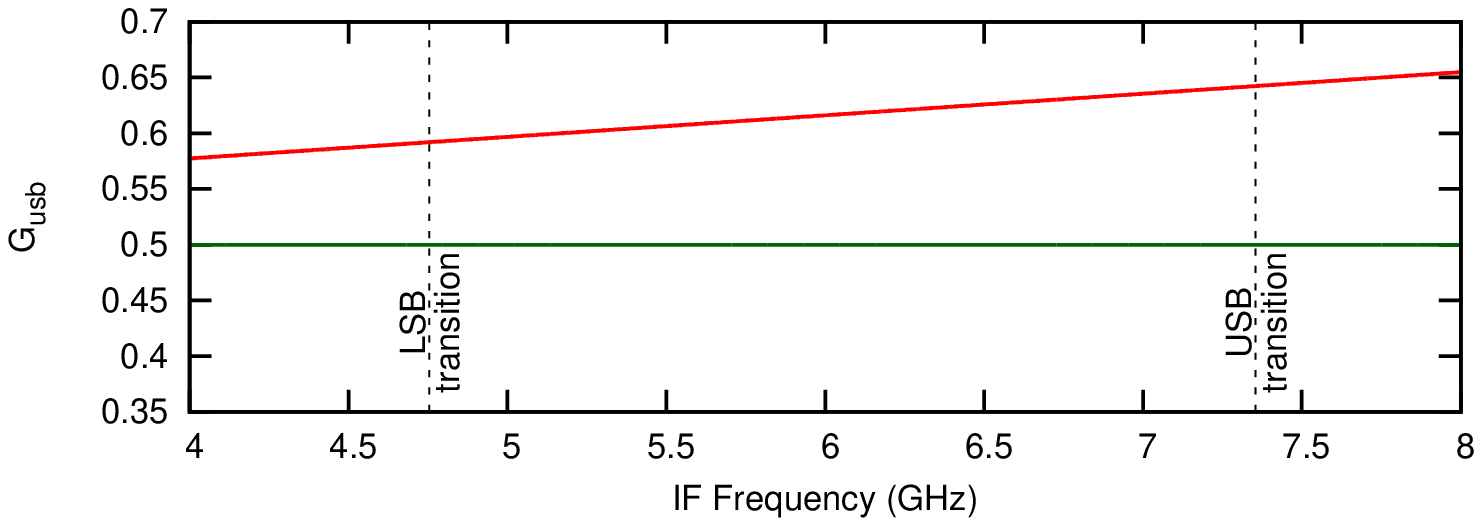}
}

\subfigure[\textit{Left}: Lower sideband intensities generated by scaling double sideband data in Fig. \ref{sbr_line_examples_dsb} with $G_{lsb}$=(1-$G_{usb}$) values shown in Fig. \ref{sbr_line_examples_gssb}. \textit{Right}: Upper sideband intensities generated by scaling with $G_{usb}$. 
Note that the left panel shows the correct amplitude only for the LSB transition and the right panel only for the USB transition.]
{
\label{glsb_correction}
\includegraphics[angle=-90.0,width=0.5\textwidth]{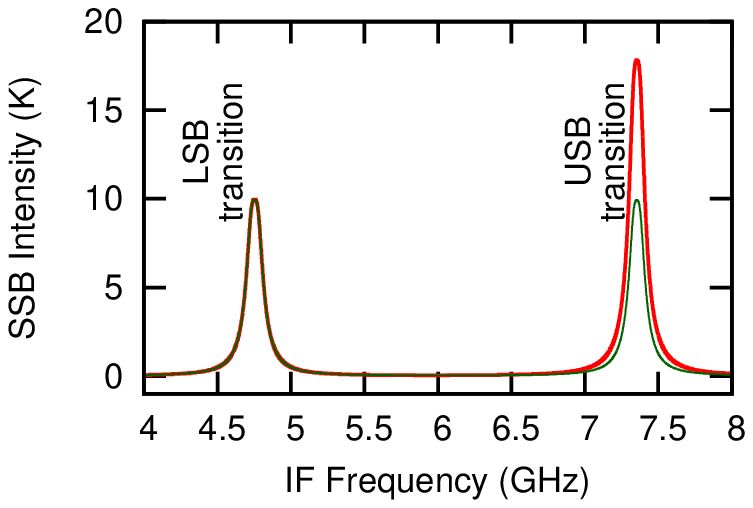}
\includegraphics[angle=-90.0,width=0.5\textwidth]{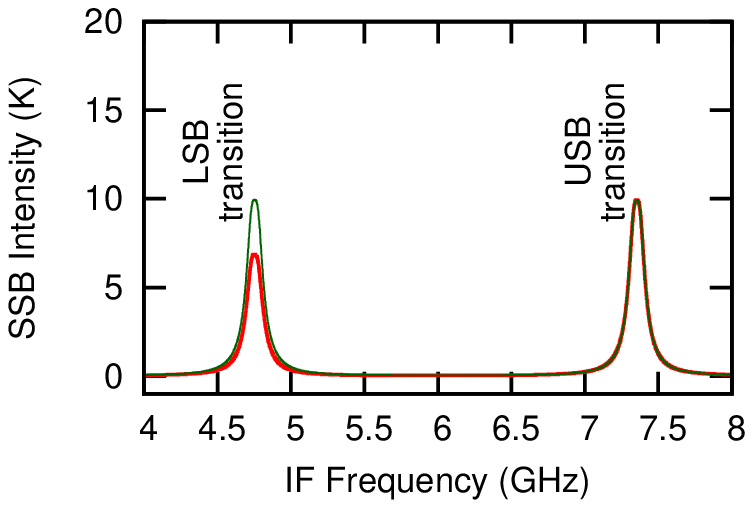}
}

\caption{The effect of a sideband imbalance on the double sideband line intensity for the molecule OCS observed at an LO frequency of 721.3 GHz. The figure highlights the need to apply the correct sideband scaling factor to the appropriate spectral line.}
\label{sbr_line_examples}
\end{figure}

\subsubsection*{Impact on the science data}

Figure \ref{sbr_line_examples} provides an overview of the heterodyne detection process and the effect of an imbalanced sideband ratio on the spectral lines. Figure \ref{sbr_line_examples_ssb_spectra} shows an example of a simulated OCS spectrum between 712 and 730 GHz with the LO tuned to 721.3\,GHz. The two OCS lines seen in the figure are at 716.546 and 728.654\,GHz which results in the down converted lines appearing at 4.75 and 7.354\,GHz respectively in the IF band. The resulting IF spectra are shown in Fig. \ref{sbr_line_examples_dsb} for the USB-dominated and balanced sideband ratio scenarios illustrated in Fig. \ref{sbr_line_examples_gssb}. The intensity in the down converted data are approximately half the original intensity, this is known as the double sideband intensity. This is similar to the level 1 data from the HIFI data processing pipeline\cite{avruch2013}.
\par
This difference in double sideband and single sideband intensity is a
result of the calibration process. Since the data are intensity
calibrated by observing a hot and cold load the final spectrum is a
fraction of the difference of their signals. Since the internal loads
provide broadband signals and contribute to both sidebands a nearly equal
continuum emission, a spectral line signal observed in only one
sideband is calibrated as approximatively half the intensity it has in
reality. 
\par 
Figure \ref{sbr_line_examples_ssb_spectra} shows equal intensities for the simulated OCS lines. In the down converted DSB spectrum shown in Fig. \ref{sbr_line_examples_dsb}, the USB-dominated sideband ratio example ($G_{usb} >$ 0.5) shows a noticeable difference in line intensity between the upper and lower sideband spectral lines. This is the effect of the sideband ratio imbalance. Where the sideband ratio is  0.5 the lines have equal intensities and are half the original SSB intensities. This example also emphasises how the SBR can noticeably vary over the IF bandwidth when a large imbalance exists in the mixer response on short frequency scales. 
\par 
Figure \ref{glsb_correction} shows an example of simulated level 2 data from the HIFI pipeline with the sideband gain ratio applied. This example highlights the importance of applying the LSB or USB sideband ratio to the respective LSB or USB line(s). When the appropriate sideband ratio is applied to the appropriate line, the single sideband line intensity can be fully recovered. Without this \textit{a priori} sideband ratio knowledge assigning the correct line intensity would not be possible.

\subsection{Diplexer effect on the sideband ratio}
\label{diplexer_mistuning_effects}

\begin{figure}
\subfigure[\textit{Left:} Correctly tuned diplexer coupling. The solid line shows the LO signal coupling, the dashed line shows the sky signal coupling. The red and blue thick lines indicate the IF bandwidth of the respective LSB and USB. \textit{Right:} Sideband ratio for a correctly tuned diplexer.]{
\label{tuned_diplexer}
\includegraphics[angle=-90.0,width=0.5\textwidth]{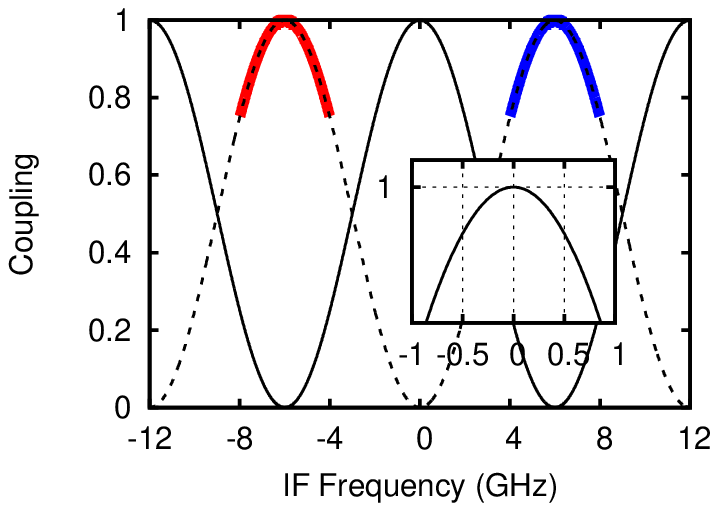}
\includegraphics[angle=-90.0,width=0.5\textwidth]{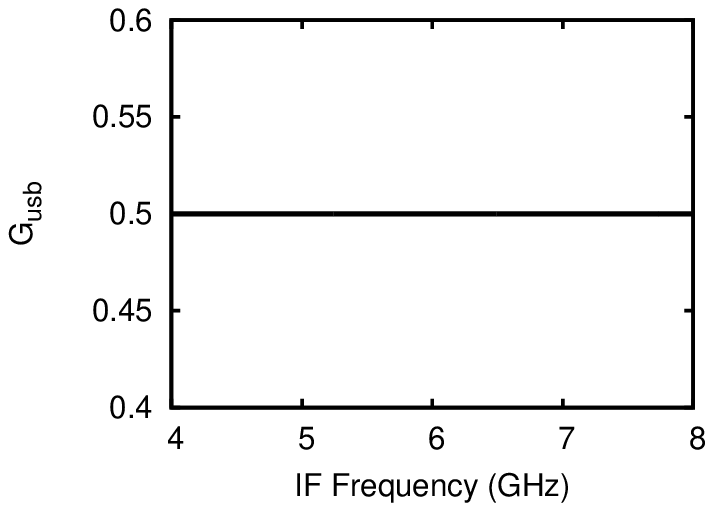}

}

\subfigure[\textit{Left:} Diplexer coupling for a -10$\mu$m mistune. \textit{Right:} Sideband ratio for a -10$\mu$m mistuned diplexer showing the IF-dependent slope.]{
\label{mis_tuned_diplexer}
\includegraphics[angle=-90.0,width=0.5\textwidth]{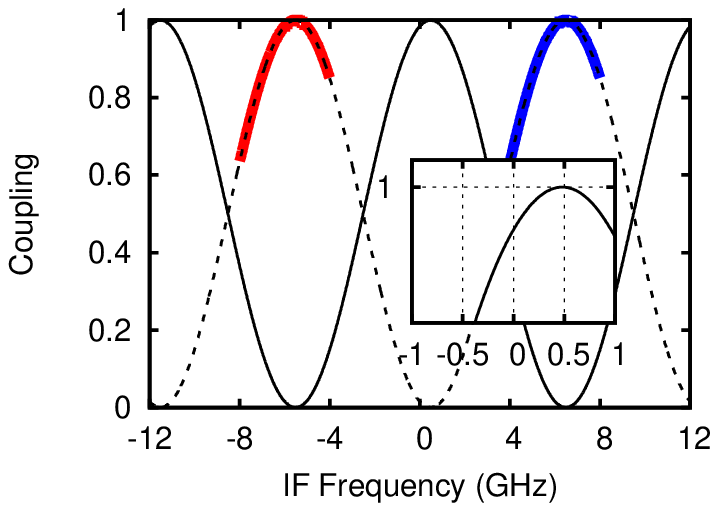}
\includegraphics[angle=-90.0,width=0.5\textwidth]{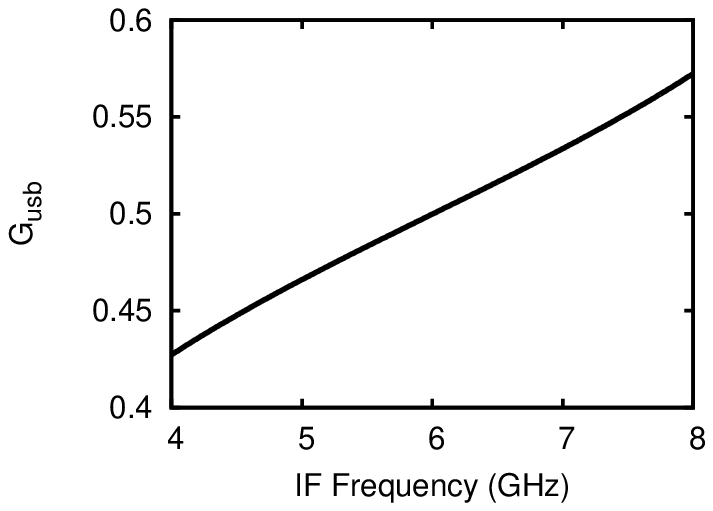}
}

\subfigure[$^{12}$CO (8-7) gas cell test data measured in the lower sideband at multiple LO frequencies showing evidence of a -10$\mu$m mistuned diplexer. Since the line is measured in the lower sideband the applicable sideband ratio is one minus that shown in Fig. \ref{mis_tuned_diplexer}.]{
\label{mis_tuned_diplexer_12co}
\includegraphics[angle=-90.0,width=\textwidth]{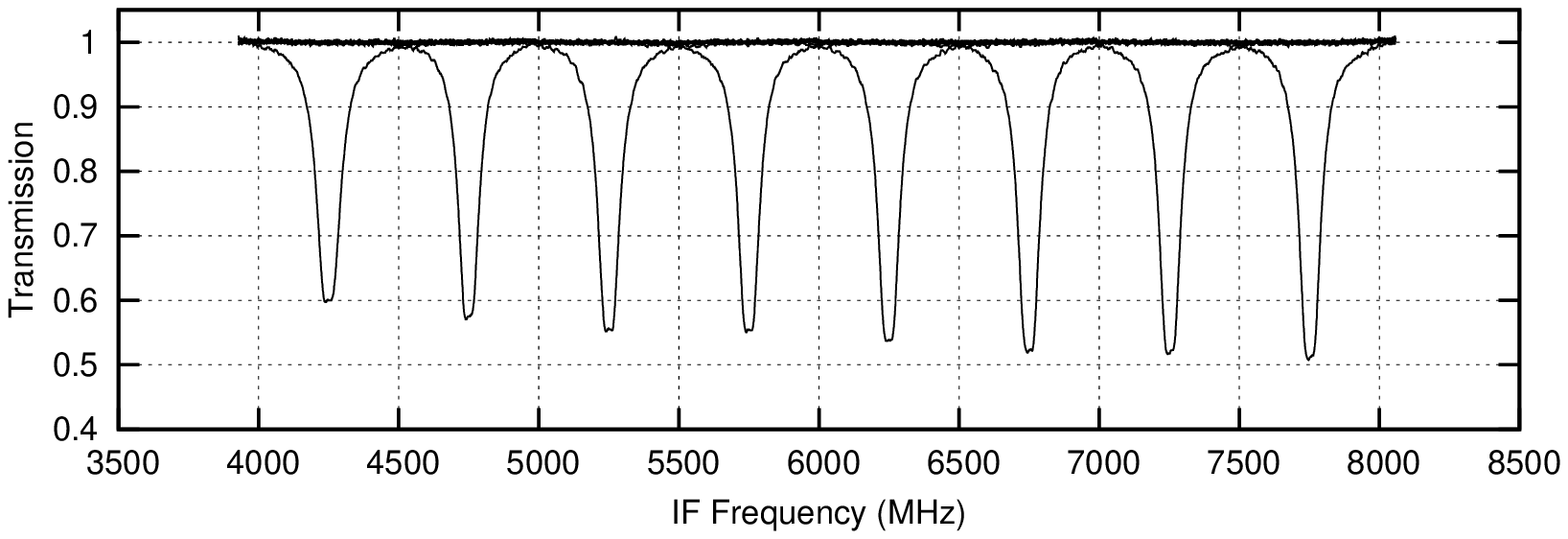}
}
\caption{The effect of diplexer mistune on the sideband ratio.}
\label{diplexer_mistuning}
\end{figure}

The LO and sky beam coupling for bands 3, 4, 6 and 7 was achieved through a Martin-Puplett diplexer (bands 1, 2 and 5 used a beamsplitter approach). The advantage of the Martin-Puplett diplexer is that the amount of LO power reaching the mixer at the correct polarization angle can be maximised. 
%The principle behind the diplexer is that the incoming LO and sky signals are split into orthogonal polarizations, the two split signals are reflected on two roof top mirrors, one fixed and one variable, by moving the variable mirror the polarization angle LO and sky signal can be rotated to match the mixer polarization angle.
%
%Simplify here - ref to Migo's paper
%The fixed and movable mirrors are offset from each other with the following length, $L$, where
%\begin{equation}
%L = \frac{c}{4\nu_{\rm if}}.
%\label{diplexer_equation}
%\end{equation}
%where $c$ is the speed of light and $\nu_{\rm IF}$ corresponds to IF band center frequency. Hence for the SIS mixer with a central IF frequency of 6 GHz there is a 20.8\,mm offset between the 2 mirrors while for the HEB mixer there is a 12.5\,mm offset corresponding to a 3.6\,GHz central frequency. By varying the movable mirror the polarization of the sky and LO signals are rotated to match the mixer polarization angle. 
%
Each LO frequency has a unique diplexer position that has the maximum LO path transmission at the LO frequency and the maximum sky path transmission peaking at the IF bandwidth centre, see Fig. \ref{tuned_diplexer}. The process of setting the diplexer position is described in more detail in \cite{mueller2013}.
\par
The diplexer can have an influence on the sideband ratio if the optimum position is not used for a given LO frequency. Ideally the respective sideband coupling through the diplexer should be symmetric, see Fig. \ref{tuned_diplexer}; however, a slight imbalance can lead to a sideband ratio slope occurring across the IF bandwidth, most prominently towards band edges, see Fig. \ref{mis_tuned_diplexer}. Fig. \ref{mis_tuned_diplexer_12co} shows an example of the $^{12}$CO (8-7) transition measured during gas cell tests at different LO frequencies in the lower sideband. The slope in line intensity induced by a diplexer mistune is noticeably across the IF band.
\par 
These effects were seen in the band 3 and 4  gas cell test data, and are particularly noticeable in saturated H$_{2}$O lines where the broad and flat saturated line plateau had a significant slope indicating a diplexer mistune\cite{higgins2011}. This diplexer mistune introduced an additional uncertainty into the sideband ratio derived from the gas cell tests. 
%Don't understand this sentence, here weg.
%Furthermore since the diplexer model was updated to account for space craft flight conditions, the sideband ratios are necessarily comparable.
Fortunately these diplexer tuning effects have not been observed in flight, as a new and improved diplexer tuning model was generated based on the ground testing experience. Investigation of $^{12}$CO (8-7) transition in orbit shows no evidence of the variation shown in Fig. \ref{mis_tuned_diplexer_12co}. However, further data mining is required for a firm confirmation that we have no diplexer induced sideband ratio effect due to mistuning.
\par 
In addition to the diplexer mistuning possibilities, the diplexer can also introduce complicated standing wave patterns. This was investigated by \citet{siebertz2007} using a band 2 prototype mixer. They showed that there can be significant standing waves issues at the IF band edges, which modulate the line intensity. \citet{delforge2013} have underpinned the experimental work with a theoretical frame work. Due to the coarse frequency sampling of the standard spectral scan used during the HIFI observations, it is difficult to measure the influence of standing waves. However, a number of dedicated tests were undertaken towards the end of the mission in order to measure the influence of this effect. These tests involved placing a spectral line towards the IF band edges and observing the line at different LO frequencies separated by tens of MHz and different diplexer mistunings. These tests will help established the degree of scatter due to standing waves and diplexer mistuning. 

\subsection{LO spectral purity}
\label{lo_purity}
It has been demonstrated that some frequency areas over the HIFI tuning range are affected by spectral impurity. What this means is that the DSB spectrum obtained from the mixing process may hold signal arising from frequency ranges not belonging to the intended ones. In this situation, the detector gain cannot be described with only 2 components (LSB and USB) but has to take into account an additional component (the Outer Sideband contribution, OSB). In practice this implies that the total system gain is altered by that of the unwanted regions and the line calibration becomes erroneous as it assumes a sideband correction relying on only two sidebands.
\par 
One can extend the formalism introduced in equations \ref{dsb_total} and \ref{g_ssb_equation} as follows. Assuming that the OSB contributes to the total gain as $\gamma_{osb}$, the new sideband gain ratio correction to apply to lines from the USB writes:
\begin{equation}
G^{impure}_{usb} = \frac{\gamma_{usb}}{\gamma_{usb} + \gamma_{lsb} + \gamma_{osb}}
\end{equation}
Since we calibrate against continuum sources (internal loads) this will manifest in a decrease of the sideband gain ratio by a factor $F_{corr}$ ($<$ 1) such that $G^{impure}_{usb} = G_{usb} \times F_{corr}$.
Consequently, one can see that $\gamma_{usb} + \gamma_{lsb} + \gamma_{osb} = (\gamma_{usb} + \gamma_{lsb})/F_{corr}$, so that the new sideband ration correction for lines in the LSB will now write:
\begin{equation}
\frac{\gamma_{lsb}}{\gamma_{usb} + \gamma_{lsb} + \gamma_{osb}} = (1 - G_{usb}) \times F_{corr}
\end{equation}
i.e. the correction factor applies similarly to lines in either sideband.
\par 
The most severe impure region was identified in band 5b, which was in fact only released to the observer community after most of its purity issues got resolved. Evidence of the sideband ratio imbalance in this band can be seen in the gas cell tests shown in Fig. \ref{gascell_SBR_overview}. In this example one can see that depending on which sideband the line is observed in, the intensity can be greatly over- or underestimated. Some other spectral ranges were purified during the mission, in particular in band 5a and in band 3b (951-953 GHz).

\section{The gas cell test campaign}
\label{gascell_sideband_ratio}
\begin{figure}[p!]
\subfigure
{
\label{sbr_bands12}
\includegraphics[angle=-90,width=\textwidth]{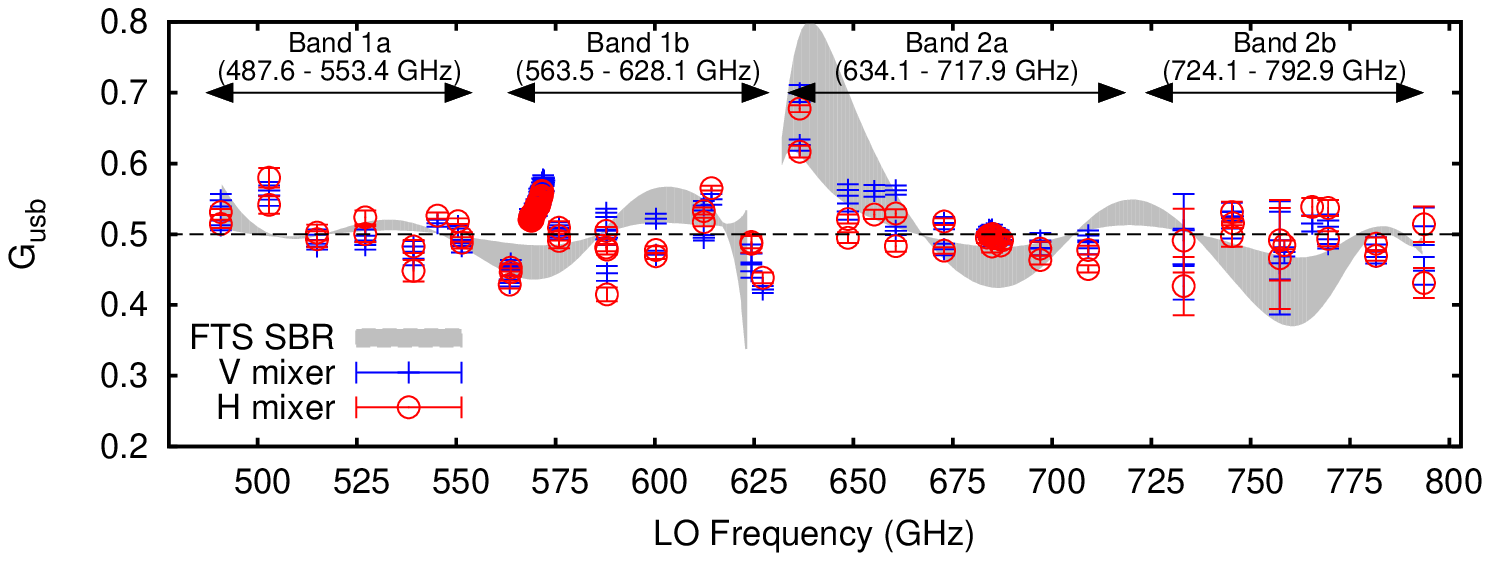}
}
\subfigure
{
\label{sbr_bands34}
\includegraphics[angle=-90,width=\textwidth]{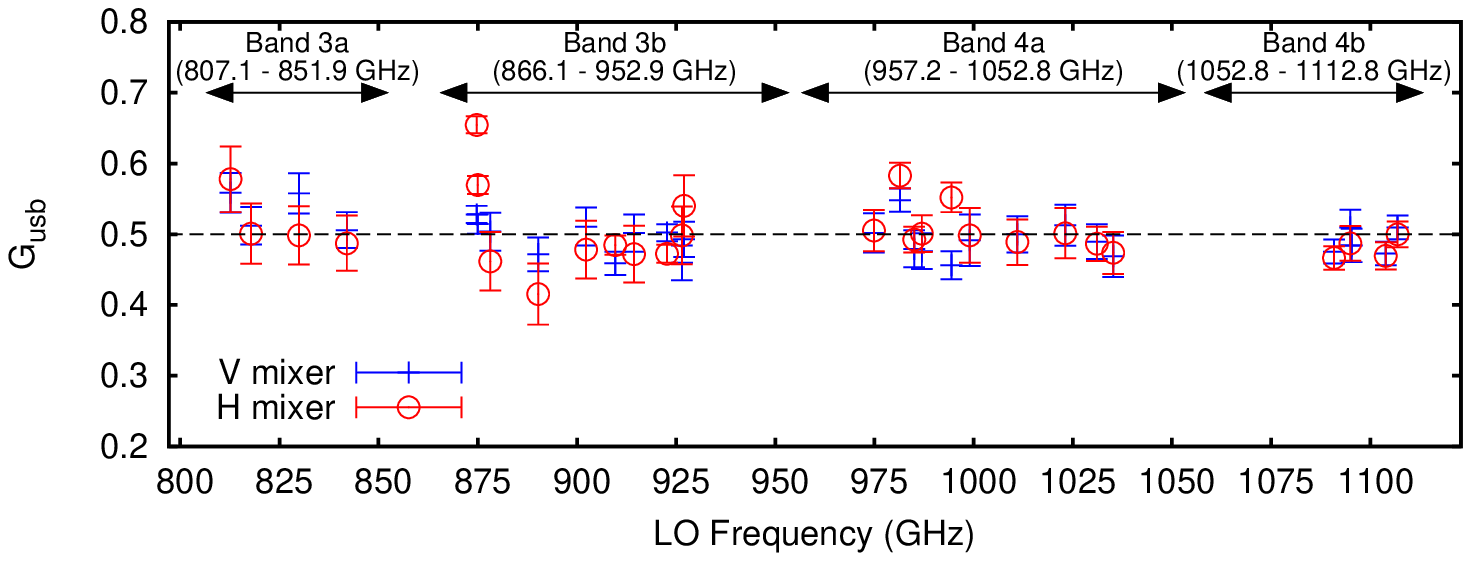}
}
\subfigure
{
\label{sbr_bands5}
\includegraphics[angle=-90,width=\textwidth]{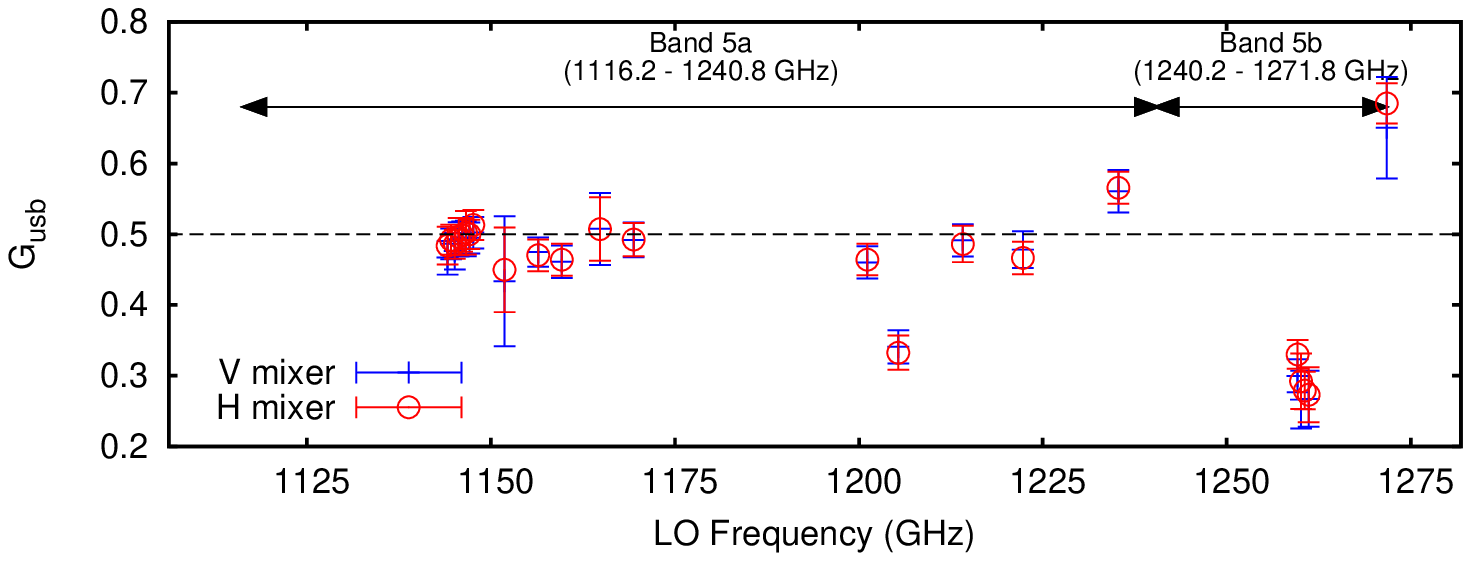}
}
\subfigure
{
\label{sbr_bands67}
\includegraphics[angle=-90,width=\textwidth]{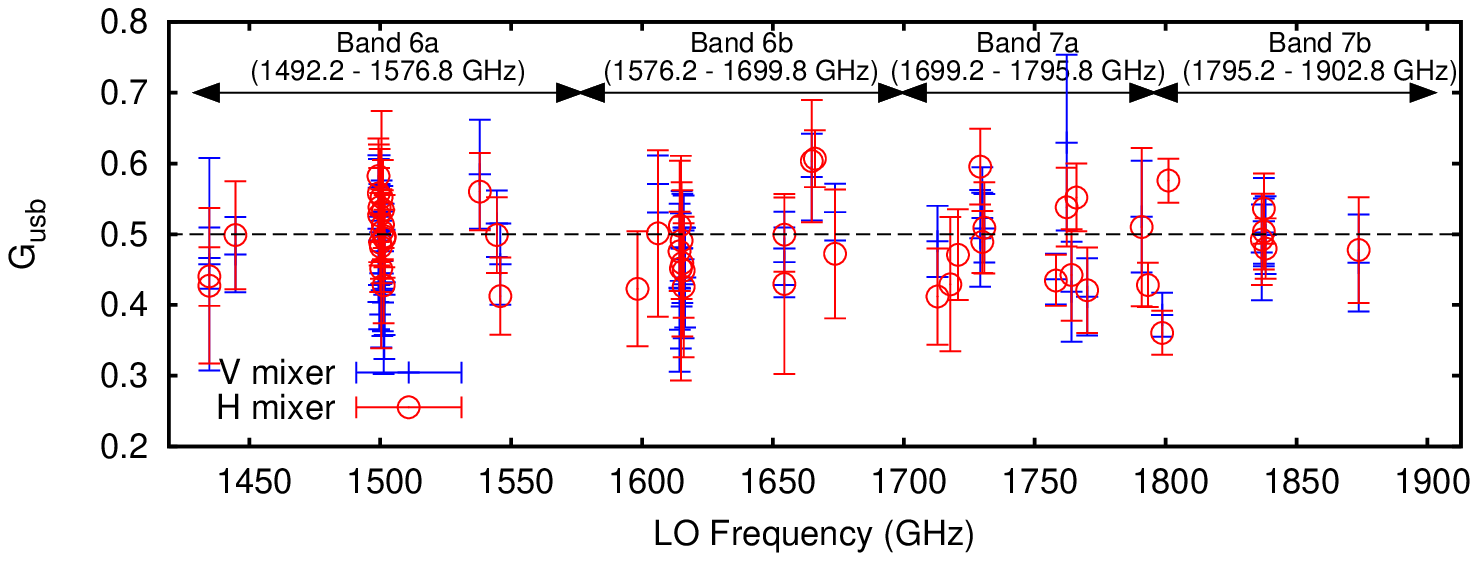}
}
\caption{Overview of sideband ratio for all H and V mixers measured using the gas cell test setup \cite{higgins2011}. A derived sideband ratio extracted from FTS measurement in Fig. \ref{fts_coupling} is plotted  in gray for bands 1 and 2. The gray area shows the maximum and minimum normalized sideband ratio calculated.
\label{gascell_SBR_overview}}
\end{figure}

Dedicated gas cell measurements were performed prior to launch to measure the HIFI sideband ratio. The test apparatus is discussed in detail in \cite{teyssier2004}, while the data acquisition and process of extracting the sideband ratio from the gas cell test data is detailed in \cite{higgins2010, higgins2011}. The HIFI gas cell was designed to present a saturated column of $^{12}$CO, $^{13}$CO, OCS and H$_{2}$O gas to the instrument detectors. By measuring these gases at various LO frequencies a sparsely sampled picture of the sideband ratio was generated. Fig. \ref{gascell_SBR_overview} gives an overview of the sideband ratios derived from the gas cell tests. 
%In the following sections we will examine some regions of interest shown in Fig. \ref{gascell_SBR_overview} using examples of science data taken during the Herschel mission.
\par 
It is interesting to compare the FTS measurements shown in Fig. \ref{fts_coupling} with the sideband ratio measured during the gas cell tests. As shown in Fig. \ref{fts_sbr_example} it is possible to generate a sideband ratio from the FTS test data. There are, however, several limitations to this exercise: while the band 3 and 4 FTS measurements were affected by the FTS setup problems mentioned in Section \ref{fts}, the gas cell test data shown for band 5 was affected by an impure LO signal (discussed in Section \ref{lo_purity}). Furthermore, the large scatter in the gas cell test data seen in bands 6 and 7 is due to poor data quality during the ground test campaign, see Section \ref{sect_heb} and \cite{higgins2009} for further details. This leaves bands 1 and 2 with the best combination of good FTS and gas cell test data.
\par 
The calculated sideband ratio extracted from the FTS measurements for bands 1 and 2 is plotted in gray behind the gas cell test results in Fig. \ref{gascell_SBR_overview}. The gray region maps out the maximum and minimum normalized sideband ratio across the IF bandwidth for each LO tuning. 
The plots shown in Fig. \ref{fts_sbr_slices} would correspond to a vertical slice of the gray FTS measurement region. 
From the plots there is some agreement between the FTS measurement and the sideband ratio inferred from the gas cell tests, in particular in the lower end of band 2a and upper end of band 1b, i.e. those areas mapping the expected drop in the mixer response towards the band edge. However, the regions associated with both $^{12}$CO transitions at 576 and 691\,GHz cannot be reconciled in the two datasets.
\par 
In the following sections we show that several of the regions of sideband ratio imbalance predicted by the gas cell tests are also seen in the flight data, which leads to the conclusion that the gas cell measurements should be more representative of the actual sideband ratio affecting the science data. We recall that the FTS measurements are performed with an un-pumped mixer, so there could be additional mechanisms other than what the FTS broadband coupling measurement can reveal, such as an LO frequency dependent heterodyne IF conversion efficiency. Some regions of sideband ratio indeed show extreme effects that contradict the assumptions made so far in this paper. The $^{12}$CO (5-4) region in band 1 for example (see Section \ref{12co_54}) shows a sideband ratio that increases towards lower IF's in both sidebands, which is opposite to the behaviour shown in the leftmost example of Fig. \ref{fts_sbr_slices}. 
%\par 
%Further research is needed to understand the relationship between FTS measured sideband ratio and %heterodyne sideband ratio. The FTS measurement is a common test used in the heterodyne community %to probe the mixer sky coupling. 

\section{Sideband ratios measured in gas cell and science data}
\label{side_band_ratio_in_data}

%As shown in \ref{sbr_line_examples} it is only possible with an aproiri knowledge of the sideband ratio to correctly assign a single sideband intensity from the measured double sideband intensity. In this section the gas cell garnered sideband ratios are presented. Following the gas cell section, a number of examples of flight data with evidence of sideband ratio imbalance will be shown. Observation methods developed by the HIFI calibration team to probe a spectral region for sideband ratio imbalance will also be presented.

\subsection{Band 1: $^{12}$CO (5-4)}

\label{12co_54}
\begin{figure}
\subfigure[Sideband ratio for the H and V mixers in the frequency range of the $^{12}$CO (5-4) and H$_{2}$O 557\,GHz transitions. The LO frequency range from 580--584 GHz covers the $^{12}$CO line as seen from the lower sideband, but it was not measured during the gas cell tests. The data shown are extrapolated from the upper sideband line using flight data to determine the relative trend.]{
\label{12co_54_gascell}
\includegraphics[angle=-90.0,width=\textwidth]{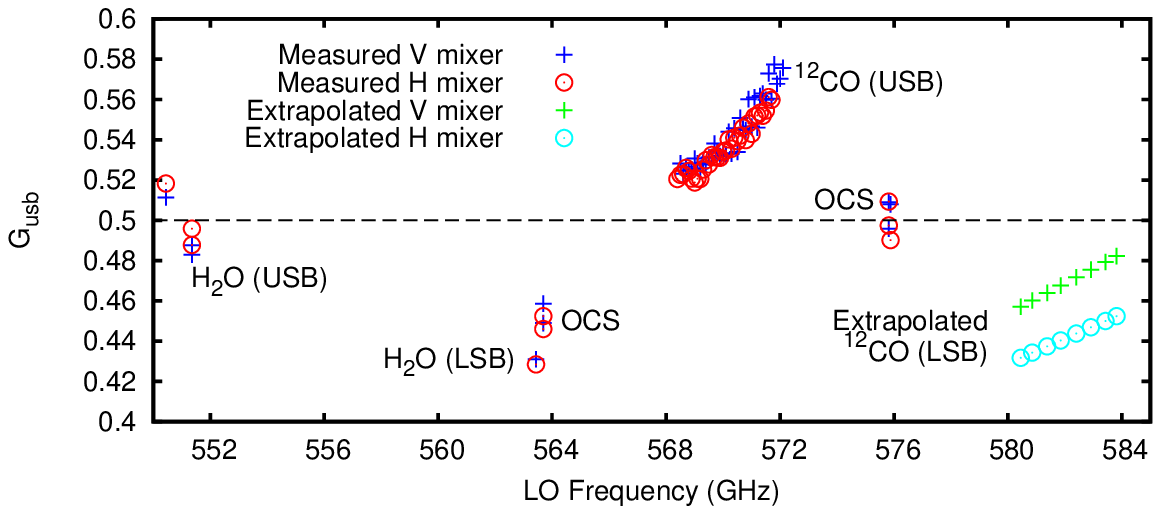}
}

\begin{comment}
$
\begin{array}{cc}
\subfigure[ $^{12}$CO (5-4) line measured with the V mixer on OMC-2 FIR 4 at different LO frequencies covering the LSB and USB showing a $\sim$10$\%$ peak intensity scatter ]{
\label{chess_line_int}
\includegraphics[angle=-90.0,width=0.5\textwidth]{chess_example_12co_WBS-V.ps}}
&
\subfigure[H and V line peak intensity versus IF for lines shown in Fig. \ref{chess_line_int}. The dashed lines show the peak intensity of the H and V deconvolved spectral lines.]{
\label{chess_line_ifplot_int}
\includegraphics[angle=-90.0,width=0.5\textwidth]{1342191591_level2_linecrop_576.246_GHz.ps}}
\end{array}
$
\end{comment}
\subfigure[\textit{Left:}$^{12}$CO (5-4) line measured with the V mixer on OMC-2 FIR 4 at different LO frequencies covering the LSB and USB showing a $\sim$10$\%$ peak intensity scatter. \textit{Right:} H and V line peak intensity versus IF for lines shown in the left panel. The dashed lines show the peak intensity of the H and V deconvolved spectral lines. The DSB line intensity was converted to SSB assuming $G_{usb}$=$G_{lsb}$=0.5.]{	
\label{chess_line_int}
\includegraphics[angle=-90.0,width=0.5\textwidth]{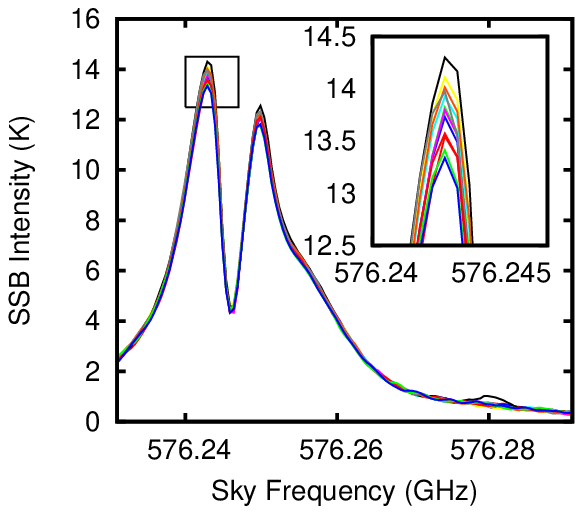}
\includegraphics[angle=-90.0,width=0.5\textwidth]{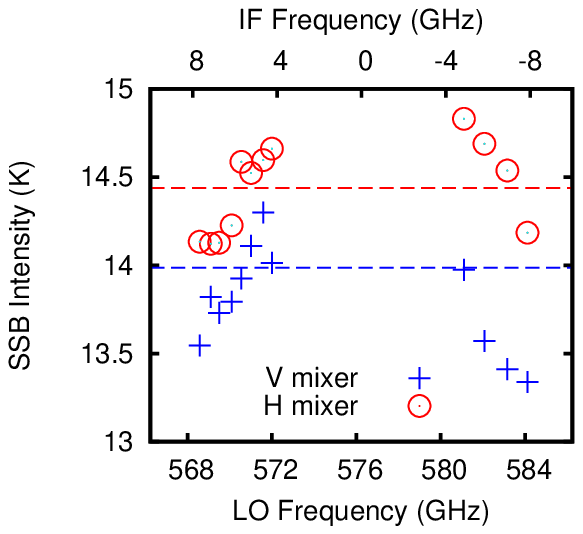}
}
\caption{
A comparison between the sideband ratio measured from gas cell tests and line intensity trends with LO frequency extracted from in flight spectral scans for the $^{12}$CO (5-4) region. Similar trends are seen in both datasets.}
\label{12co_54_plot}
\end{figure}

Figure \ref{12co_54_gascell} shows the sideband ratio extracted from gas cell test data between LO frequencies of 550 and 585 GHz. Noteworthy here is the similarity in sideband ratio between the H and V mixer. In this plot we see the sideband ratio extracted using three different molecules, H$_{2}$O, $^{12}$CO and OCS. From the figures one can see that depending on which LO frequency the $^{12}$CO line is measured at there is a noticeable change in sideband ratio and therefore line intensity. Unfortunately the $^{12}$CO (5-4) transition was only measured in the upper sideband in the gas cell tests. We can, however, use flight data from the upper and lower sideband and extrapolate the sideband ratio measured during the gas cell tests from the upper sideband into the lower sideband. 
\par 
Figure \ref{chess_line_int} shows an example of the $^{12}$CO line from OMC-2 FIR 4 taken as part of the CHESS key program\cite{kama2013}. The figure shows multiple observations of the $^{12}$CO line at different LO frequencies and reveals a $\sim$10$\%$ scatter in line intensity. When these intensities are plotted as a function of LO frequency the variation in the upper sideband trend is consistent with the trend measured in the gas cell test data. Using the spectral survey data from the upper sideband the intensity variation into the lower sideband can be examined. The data from Fig. \ref{chess_line_int} indeed shows that there is a similar IF-dependent trend in the lower sideband as the line intensity is seen to increase towards lower IF (points at negative IF). Using the gas cell measurements of the upper sideband behaviour of the $^{12}$CO line in combination with the spectral scan data it is possible to extrapolate the sideband ratio into the region where the $^{12}$CO transition is in the lower sideband (LO frequencies between 580-584GHz). The extrapolation is undertaken by correcting the upper sideband line with the correct sideband ratio. In the example shown the sideband corrected peak line intensity is estimated to be 13.5K for the H mixer and 13K for the V mixer. Using this intensity as a reference, the lower sideband ratio in the 580-584 GHz can be derived. The extrapolated sideband ratios are shown in Fig. \ref{12co_54_gascell}. From the plot it is apparent that an abrupt turnover in sideband ratio is required in order to match the upper sideband intensity to the lower sideband one. This turnover is supported by the OCS observation at an LO frequency of $\sim$576 GHz which suggests a nearly balanced sideband ratio of 0.5 at an LO frequency between the two sidebands where the $^{12}$CO transition is measured.

%A typical processing step for spectral scan data is the deconvolution algorithm\cite{comito2002}. By note the movement of spectral line with changing LO frequency it is possible to disentangle the a series of double sideband spectra at different LO frequencies into a single sideband spectra. The deconvolution becomes more difficult when the sideband ratio problem is included. The dashed line in figure \ref{chess_line_ifplot_int} shows the peak intensity of the deconvolved spectral line when the sidebands are assumed to have equal gain. For key programs such as CHESS this deconvolved spectra is the main science product. In this $^{12}$CO example the deconvolved line intensity would be $~\pm$ over estimated. 

\par 
The $^{12}$CO region we have discussed here serves to illustrate the complexity of the sideband ratio problem. Fig. \ref{12co_54_gascell} evidences that 
there is a certain level of structure that is not accurately probed between the spot frequencies collected in the gas cell test campaign. The first approach to the gas cell test results was to simply fit a polynomial through the measured sideband ratio and then interpolate between the measured points. The example shown here illustrates the futility of such an approach. It was initially assumed, based on the FTS measurements, that the sideband ratio would be a smooth function across the band; however, this is not the case everywhere. Further investigation is needed to explain the sharp features that are seen in the data. The next section examines another region of band 1 which is of particular importance to the HIFI science goals and shows a similar sharp sideband ratio feature.

\subsection{Band 1: H$_{2}$O at 557\,GHz}
\label{h2o_band1a_1b}
The observation of water was one of the main science goals of
HIFI. The ground state water lines occur around 557 GHz ({\it ortho}-H$_2$O, band 1) and 1.11 THz
({\it para}--H$_2$O, band 4). The 557\,GHz water line covers the
boundary between the upper and lower LO bands of the band 1 mixer. The
lower region is known as LO band 1a and observes the line in the USB, while the upper region is the LO band 1b and observes the line in the LSB. From the observation of a single water line alone various properties can be extracted; however, interestingly from the ratio of the {\it ortho}-to-{\it para} line intensity the physical conditions when the gas was formed can be determined\cite{lis2010}. In this section we will discuss the {\it ortho}-H$_2$O line in band 1. 
%The para transition is discussed in the context of diplexer mistuning in section \ref{diplexer_mistuning_effects}.

\begin{figure}[t!]
\includegraphics[angle=-90.0,width=\textwidth]{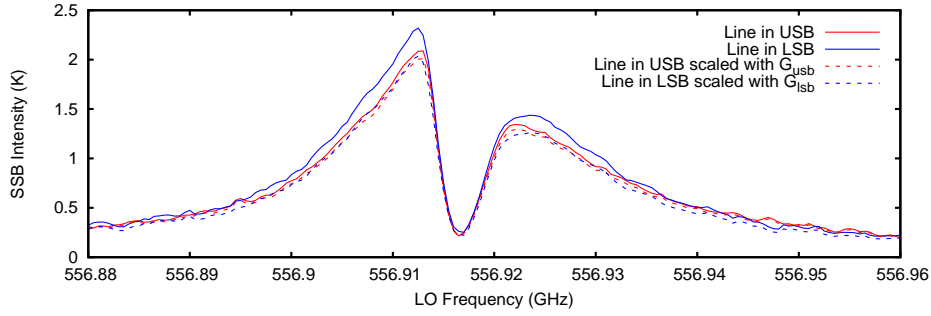}
\caption{Sideband ratio imbalance seen in 557\,GHz {\it o}-H$_{2}$O line from OMC-2 FIR 4 \cite{kama2013}. The deconvolved line profile from band 1a (red) and 1b (blue) is shown here. The line before sideband ratio correction is shown with the solid line, the dashed line shows the corrected line profile.}
\label{h2o_ortho_plot}
\end{figure}

\par 
%Due to this coincidence a number of key programs observed water only in one sideband and hence don't see some of the discrepancies that exists between the upper and lower sideband line intensity which hint to underlying sideband ratio effects.
Figure \ref{h2o_ortho_plot} shows an example of the deconvolved 557 GHz line observed in both upper and lower sidebands using the H mixer as part of the CHESS key program. The line observed in the USB is 10$\%$ stronger than that observed in the LSB. This is consistent with the sideband ratio measured during the gas cell tests shown in Fig. \ref{12co_54_gascell}. When the sideband ratio derived from the gas cell tests is applied to the respective USB and LSB measured lines, the discrepancy between the USB and LSB line intensity is greatly reduced. Much like the $^{12}$CO region discussed in the previous section this extreme change in sideband ratio is not predicted by the FTS measurements and requires a rather extreme variation in broadband coupling to reproduce such an effect.
\par 
What is interesting in this example is that the line intensity is overestimated both when it is the upper and lower sideband. When the appropriate sideband ratio is applied to the respective sideband measurements, the line intensity is reduced for both upper and lower sideband lines (Fig. \ref{h2o_ortho_plot}). Without knowledge of the sideband ratio, a standard approach here would be to average the two line profiles together which would result in an overestimated line intensity.
\par 
For further information on the application of sideband ratios after the pipeline processing see reference \cite{teyssier2013}. This technical note contains a table of sideband ratios at spot frequencies and describes how to apply the sideband ratio to level 2.

\subsection{Band 2 : lower band edge}
\label{band2a_lower_edge}
\begin{figure}
\includegraphics[angle=-90.0,width=\textwidth]{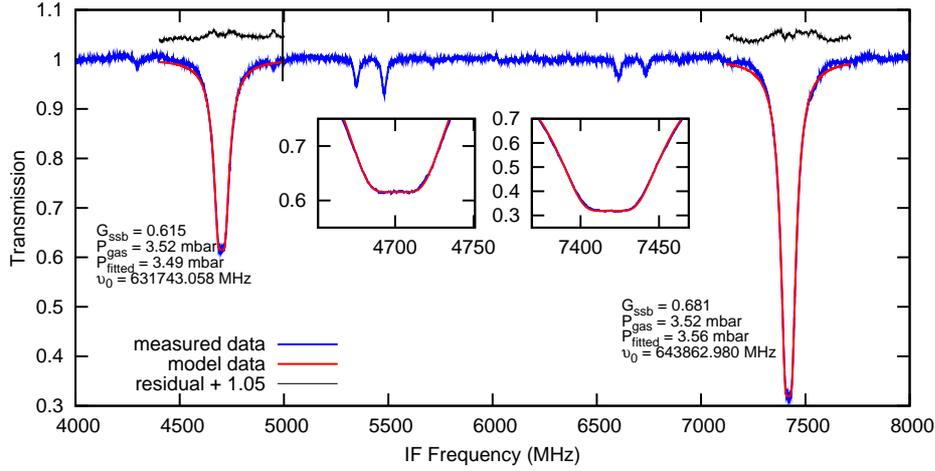}
\caption{Gas cell measurement of the $J$=52-51 and $J$=53-52 transitions of OCS in the HIFI Wide-Band-Spectrometer H at LO = 636.444 GHz. The red profiles correspond to the best LTE (Local Thermal Equilibrium) fits -- note that both lines are predicted to be saturated. The SBR at the IF centre is expected to be $G_{usb}$ = 0.652; however, the values derived from the respective transitions at the two IF borders differ from this due to the additional IF-dependent slope.}
\label{ocs_band2a}
\end{figure}

One of the most noticeable sideband ratio variations can be observed at the lower end of band 2a. In this range, the mixer response is dropping steeply downwards of 650\,GHz and the imbalance between the two sideband gains increases accordingly, leading to an upper sideband being more efficient than the lower sideband. This mixer response drop is well measured by the gas cell test data obtained in this frequency range, as can be seen in Fig.~\ref{sbr_bands12}. The grey area in this plot also shows that the trend is consistent with the FTS broadband coupling measurement\cite{teipen2005}. This strong imbalance is illustrated in Fig. \ref{ocs_band2a}, showing a gas cell measurement of two OCS subsequent transitions present in the respective sidebands at an LO frequency of 636.4\,GHz. In a sideband-balanced scenario the OCS transitions should have similar line intensities, however the line from the LSB exhibits a much weaker absorption than its USB counterpart (note that both lines are saturated). This measurement also demonstrates that, in regions of steep mixer gain, the sideband ratio also has a noticeable IF-dependency, scaling typically with the ratio applying at the centre of the IF (see also Appendix A in \cite{roelfsema2012}). The spectrum measured with the gas cell provide a real world example of the simulation described in Fig. \ref{sbr_line_examples}.

%%Caption of Fig. "OCS_2a.tiff": Gas-cell observation of the $J$=52-51 and $J$=53-52 transitions of OCS in the WBS-H at LO = 636.444 GHz. The red profiles correspond to the best LTE fits – note that both lines are predicted to be saturated. The SBR at the IF centre is expected to be $G_{ssb}$ = 0.652, however the values derived from the respective transitions at the two IF borders differ from this due to the additional IF-dependent slope.

\subsection{Band 5}
\label{band_5_slope}
\begin{figure}
\includegraphics[width=\textwidth]{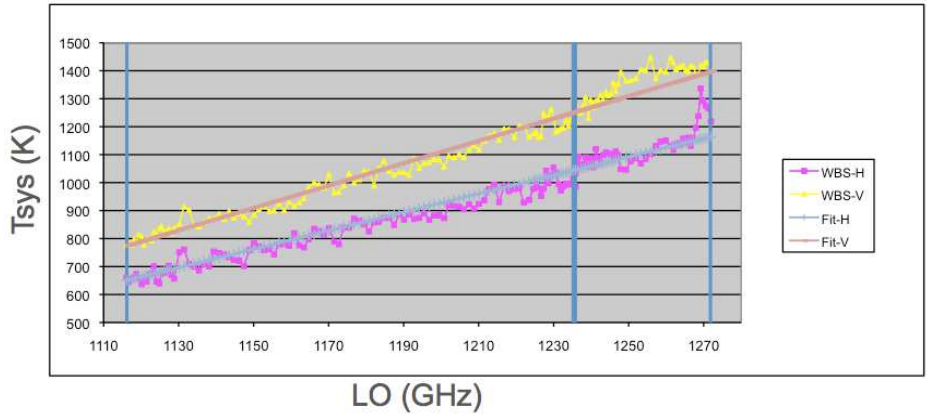}
\caption{System noise temperature (DSB) for the H and V polarisations of the band 5 mixers. The vertical bars indicate the ranges for the LO band 5a and 5b respectively. The straight lines correspond to a constant slope of 6\%.}
\label{band5_slope}
\end{figure}

In the mixers used for the HIFI band 5, the coupling between the antenna and the SIS devices is effectively a resistor (normal metal) where there is a slow but increasing loss with frequency. This response characteristic is reflected in the system noise temperature of the mixer, whereby a constant slope is observed over the whole operational range, see Fig. \ref{band5_slope}. We have interpreted this behaviour as a systematic loss in the RF (Radio Frequency) input of the USB relative to that of the LSB. With an estimated slope around 6\%, the corresponding normalised sideband ratio for a line in the USB is $G_{usb} = 0.485$.
\par 
While the limited number of points from the gas cell test data in band 5 suggests a fairly constant sideband ratio, potentially below 0.5 within the error, checks on selected strong lines observed in orbit in both sidebands confirm this trend overall.
\par 

\subsection{Bands 6 $\&$ 7}

%\textcolor{red}{\textbf{shows examples much like the h2o and 12co lines where we see no scatter and everything looking rosy.}}
As summarized in Table \ref{mixer_overview}, the HEB mixers in bands 6 and 7 have a smaller bandwidth of 2.4\,GHz, centered at 3.6\,GHz, compared with the SIS mixer of bands 1--5. From Fig. \ref{fts_sbr_example} it is apparent that lower IF's are less sensitive to sideband ratio imbalances than higher ones. Furthermore the HEB mixer unit has an inherently flatter response due to the mixer and double slot antenna technology used (similar antennae are used in band 5). This is also noticeable from the FTS measurements in Fig. \ref{fts_coupling} which shows a slowly varying broadband coupling across the mixer range. 
\par 
Nevertheless, the sideband ratio measured in the HEB bands during the gas cell tests, and summarized in
Fig. \ref{gascell_SBR_overview}, show a large degree of scatter. In the original HIFI calibration paper \cite{roelfsema2012}, the sideband ratio error was taken as the standard deviation of the sideband ratios across a given band, which has a large impact on the overall calibration accuracy of the HEB bands. However, when comparing flight data to gas cell test data, we see none of this scatter.
\begin{figure}
$
\begin{array}{cc}
\subfigure[Band 7a]{
\label{heb_line_1}
\includegraphics[angle=-90.0,width=0.5\textwidth]{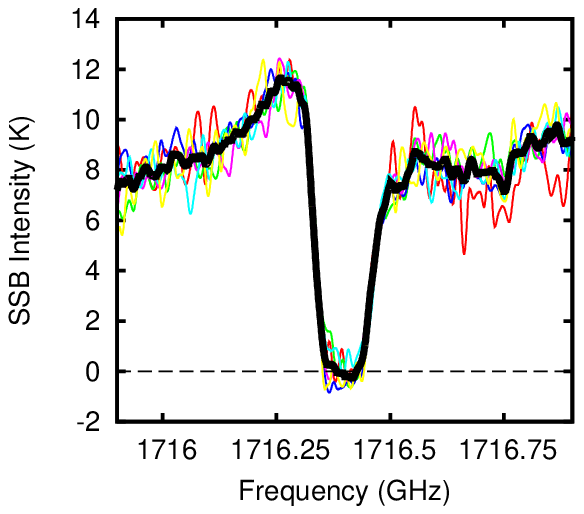}}
&
\subfigure[Band 6b]{
\label{heb_line_2}
\includegraphics[angle=-90.0,width=0.5\textwidth]{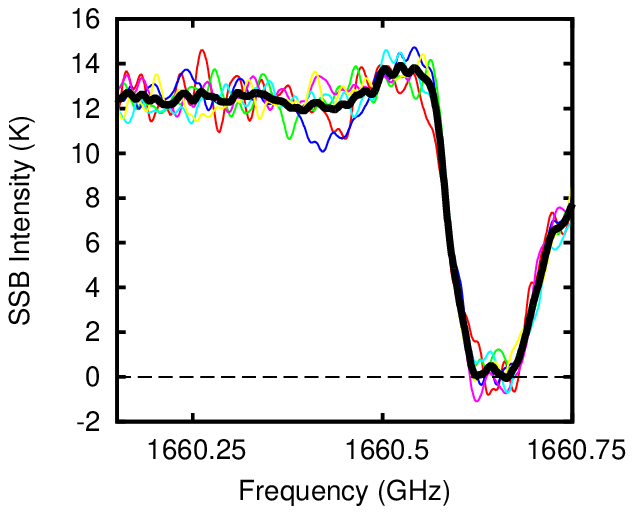}}
\end{array}
$
\caption{Saturated absorption lines seen in the H polarization towards SgrB2(M) in bands 7a and 6b. The dark line is the average of more than 8 line profiles measured at different IF's in both the upper and lower sidebands.}
\label{heb_line_example}
\end{figure}
\par 
Figure \ref{heb_line_example}  shows an example of two saturated
absorption lines seen towards SgrB2(M), near the Galactic Centre. In this example
a foreground cool gas cloud absorbs all the continuum background radiation at the observed molecular transition frequency. The spectra are plotted on a single sideband intensity scale and hence, for a fully saturated line the minimum is seen at 0\,K, indivative of a balanced sideband ratio scenario. The data shown here is taken from a spectral scan observation, so that the line is observed at different LO frequencies and hence different IF's. Spectra from both the upper and lower sidebands are overplotted and there is no discernible difference in line intensity, providing further evidence of a flat mixer gain at the scale of the IF bandwidth.
\par 
Returning to the gas cell test data, 
%upon closer inspection, supported by evidence from the flight data, 
it was clear that the experimental set up used to measure the sideband ratio in the HEB bands was not representative of the way the data were going to be taken in flight\cite{2008dieleman}. The LO unit was mounted in a separate cryostat to the main HIFI unit, resulting in a degraded system stability due to path length modulation between the two cryostats. With the reduced system stability, the data were beset with standing wave issues. Even though the gas cell test data were corrected from spectral baseline distortion, the extracted sideband ratio still showed large variations. Therefore, we believe that the sideband ratio measured during the gas cell tests in the HIFI bands 6 and 7 is not representative of the true mixer response in those bands and should be ignored for future error budgets. Further data mining will be undertaken during the post operation period to confirm that the sideband ratio across the HEB bands is 0.5. All evidence so far supports this assumption.

%obsid examples 1342216702 1763.35 GHz, 1716.4 7b  look for C+ example\newline 
%obsid examples 1342216501 1660.65 GHz 6b
\label{sect_heb}
\section{Discussion and lessons learned}
The gas cell test campaign provided thousands of measurement points prior to the launch of Herschel, building probably one of the largest dataset dedicated to the calibration of the sideband ratio in a double sideband receiver, and demonstrating the unique capabilities of HIFI as a high-resolution spectrometer already years before the Herschel launch. Yet, more than half a year after the mission completion the sideband ratio correction is still puzzling in several areas of the HIFI frequency range. One of the main reasons for this is the relative scarcity of measurement points that was inherent to the limited number of gases available for the gas cell tests. This frequency coverage granularity later proved to be insufficient in bands such as band 1 or 2 where steep sideband ratio variations have been confirmed to occur over relatively narrow frequency domains. 
\par
The best way to join the dots between the pre-launch measurement
points is to make use of high signal-to-noise spectral sweeps
that can provide relative variation of line intensities within
frequency steps of a few hundred MHz. The combination of the absolute
sideband ratio points derived from the gas cell tests, and their relative evolution
in flight data is therefore needed to derive the detailed structure of
the gain profile over the whole HIFI range. This approach was
considered before launch already when a full spectral scan of the
methanol molecule was collected in the laboratory. While this dataset
probably contains most of the answers we are after, it is to date
still too complex to interpret due to often non-mature spectroscopic
parameters (e.g. pressure broadening) for methanol, which is a
mandatory input for a correct modelling of the molecular absorption
(most of it is non-saturated) as a function of the frequency.
\par
We note that part of the complexity uncovered in e.g. band 1 was not anticipated, even during the design of the mixer circuits. From the early FTS measurements of the mixer broadband coupling, it was unclear that significant variations of the sideband ratio over frequency ranges as narrow as 4 GHz, such as the one evidenced e.g. around the $^{12}$CO 5-4 line (Fig. \ref{12co_54_plot}), would apply to the mixer response. In hindsight OCS measurements on a finer LO frequency grid would have helped unravel the coupling profile of bands 1 and 2. It is still unclear whether the discrepancy between those FTS measurements and the gas cell test data is simply due to the fact that the FTS tests did not use pumped mixers. We suggest that this discrepancy could be further investigated by groups involved in heterodyne detector design.
\par
One of the important questions for the astronomers is, what is the
final absolute accuracy of the line intensity observed, which for HIFI
often boils down to the contribution of the sideband ratio. Because
the sideband ratio is currently assumed to be constant over most of the
tuning range, the same is true for its accuracy, which is
simply taken as the standard deviation of the sideband gains derived from the gas cell tests over a given band.
We discussed in Section~\ref{sect_heb} that this is too pessimistic e.g. for the HEB
bands. On top of that, our description of the sideband ratio is LO- and IF-dependent, hence the calibration accuracy should be treated accordingly. Other aspects of the sideband ratio calibration, such as standing wave and/or diplexer cross-talk (not discussed in detail in this paper), are too complex to measure at such granularity. This is where the need of a detailed instrument model becomes important, and such work is currently on-going within the HIFI team \cite{delforge2013}. It is unrealistic to assume that a theoretical model could accurately predict the frequency-dependent sideband ratio profile at any given LO frequency, so the idea here is not to correct observational data by synthetic instrument response function. Rather, such a model should provide the order of magnitude of typical calibration errors resulting from optical effects such as standing waves, esp. in the diplexer bands.\par 
An important lesson from the sideband ratio calibration effort is that a large fraction of the answers can be directly extracted from the science data themselves. While the gas cell test campaign offered measurement conditions with a fully controlled sky signal input, it was relatively constrained in time and performed at a moment when the instrument understanding was not fully mature, so some of the data peculiarities where still uncovered or not accounted for at the time. The regular science observations gathered over the almost four years of mission provide the complementary information that not only allows to revisit some of the laboratory data, but also to probe the instrument behaviour in an often larger variety of parameter space than could be considered pre-launch. The combination of the two is the key to build the most accurate picture of the receiver characteristics, and offer the best possible calibration to the legacy archive data.\par
In that respect, the main source of information lies in the so-called spectral surveys, where a given spectral range is observed several times at nearby LO frequency tunings. This redundancy allows to invert the Double Sideband problem and build a Single Sideband solution - this is called {\it deconvolution} \cite{comito2002}. In this process, the gains applying respectively to the LSB and USB can be fitted in order to reconcile all individual DSB spectra with the SSB solution. Intrinsically the sideband gain profile inferred from this algorithm is only relative and needs to be tied to absolute values measured somewhere else (typically during the gas cell tests). The quality of the recovered sideband gain profile depends on the line density (if it is too large the risk of line blend is high, if it is too low the problem becomes degenerated at some frequencies). The HIFI team is currently running simulations on synthetic data with a variety of user-fed sideband ratio models and line density to identify the best data-set of spectral surveys to be used for that purpose, together with the accuracy associated with such an approach. The goal will be to extract an as continuous as possible profile of the sideband ratio over the HIFI observational range. Until then, the recommendation is to use the sideband ratio measured at particular line frequencies during the gas cell test campaign. This information, along with the calibration uncertainty associated with the sideband ratio, is communicated via the technical note\cite{teyssier2013} found on the HIFI calibration webpages\footnote{http://herschel.esac.esa.int/twiki/pub/Public/HifiCalibrationWeb/}.

\begin{acknowledgements}
We would like to thank Mihkel Kama from the CHESS team for his help in investigating the sideband ratio effect around the 557 GHz water line. Ronan Higgins would like to thank Netty Honingh for useful discussions on the origins of the sideband ratio imbalance in bands 1 and 2. The authors are grateful to the anonymous referee for valuable comments.
\end{acknowledgements}

\bibliographystyle{spphys}       % APS-like style for physics
%\bibliographystyle{aa}       % APS-like style for physics


%\bibliography{bibliography.bib}

\end{document}